\documentclass[aps,prl,nofootinbib,twocolumn,superscriptaddress,preprintnumbers,balancelastpage,longbibliography]{revtex4-1}

\usepackage{aas_macros}
\usepackage{titletoc}
\usepackage[normalem]{ulem}
\usepackage{amsmath,amssymb,amsfonts,graphicx}
\usepackage{dcolumn}
\usepackage{bm}
\usepackage{color}
\usepackage{bbold}
\usepackage[dvipsnames]{xcolor}
\usepackage{placeins}
\usepackage{slashed}
\usepackage{multirow}
\usepackage{subfigure}
\usepackage{tabularx}
\usepackage{mathtools}
\usepackage{blindtext}
\usepackage{soul}
\usepackage{slashed}
\usepackage{amsmath}
\usepackage{tikz}
\usepackage[compat=1.1.0]{tikz-feynman}

\usepackage{hyperref} 
\hypersetup{
    colorlinks=true,       
    linkcolor=blue,        
    citecolor=blue,        
    filecolor=magenta,     
    urlcolor=blue          
}

\definecolor{coral}{RGB}{255,127,80}
\definecolor{indigo}{RGB}{75,0,130}
\definecolor{red}{rgb}{0.9, 0,0}
\definecolor{cerulean}{rgb}{0., 0.62,0.9}
\definecolor{navy}{rgb}{0.05, 0.05,0.8}

\newcommand{\scat}{\rm sc}
\newcommand{\ann}{\rm an}

\newcommand{\TeV}{{\rm \, TeV}}
\newcommand{\GeV}{{\rm \, GeV}}
\newcommand{\MeV}{{\rm \, MeV}}
\newcommand{\keV}{{\rm \, keV}}

\begin{document}
\preprint{CERN-TH-2025-152, TTP25-024, P3H-25-052}
\title{The SN~1987A Cooling Bound on Dark Matter Absorption in Electron Targets}
\author{Claudio Andrea Manzari}
\email{camanzari@berkeley.edu}
\affiliation{Berkeley Center for Theoretical Physics, Department of Physics, University of California, Berkeley, CA 94720, USA}
\affiliation{Theoretical Physics Group, Lawrence Berkeley National Laboratory, Berkeley, CA 94720, USA}
\author{Jorge Martin Camalich}
\email{jcamalich@iac.es}
\affiliation{Instituto de Astrof\'isica de Canarias,  C/ V\'ia L\'actea, s/n E38205 - La Laguna, Tenerife, Spain}
\affiliation{Universidad de La Laguna, Departamento de Astrof\'isica, La Laguna, Tenerife, Spain}
\affiliation{CERN, Theoretical Physics Department, CH-1211 Geneva 23, Switzerland}
\author{Jonas Spinner}
\email{j.spinner@thphys.uni-heidelberg.de}
\affiliation{Institut f\"ur Theoretische Physik, Universit\"at Heidelberg, Germany}
\author{Robert Ziegler}
\email{robert.ziegler@kit.edu}
\affiliation{Institut f\"ur Theoretische Physik, Universit\"at Heidelberg, Germany}
\affiliation{Institut f\"ur Theoretische Teilchenphysik, Karlsruhe Institute of Technology, Karlsruhe, Germany}

\begin{abstract}
We present new supernova (SN~1987A) cooling bounds on sub-MeV fermionic dark matter with effective couplings to electrons. These bounds probe the parameter space relevant for direct detection experiments in which dark matter can be absorbed by the target material, showing strong complementarity with indirect searches  and constraints from dark matter overproduction. Crucially, our limits exclude the projected sensitivity regions of current and upcoming direct detection experiments. Since these conclusions are \textit{a priori} not valid for light mediators, we extend our analysis to this case. We show that sub-GeV mediators can be produced resonantly both in supernova cores and in the early Universe, altering the SN 1987A analysis for effective couplings. Still, a combination of supernova cooling constraints and limits from dark matter overproduction excludes the entire parameter space relevant for direct detection in this case.
   
\end{abstract}

\maketitle

The nature of dark matter (DM) remains elusive to this day. Direct detection (DD) experiments have
pushed the limits on elastic scattering cross sections on nucleons close to the neutrino floor for multi-GeV DM. Combined with the so-far negative searches for particles with weak-scale masses at colliders and in astrophysical indirect detection (ID), the present experimental situation motivates the exploration of alternative DM scenarios beyond the classic WIMP paradigm (see, e.g. Refs.~\cite{Cirelli:2024ssz,Arcadi:2024ukq}).

One of the prominent emerging directions involves candidates with masses below 1~GeV. For such light DM, nuclear recoil signals in DD experiments typically fall below detection thresholds, which are often of the order of $\mathcal{O}({\rm keV})$ and, as a result, sub-GeV DM remains largely unconstrained by conventional experimental techniques. This challenge has motivated the development of lower-threshold detectors and alternative scattering targets~\cite{Essig:2011nj,Graham:2012su,Essig:2012yx,Guo:2013dt,Hochberg:2015pha,Essig:2015cda,Hochberg:2015fth,Hochberg:2016ajh,Schutz:2016tid,Hochberg:2016ntt,Derenzo:2016fse,Hochberg:2016sqx,Knapen:2016cue,Essig:2017kqs,Budnik:2017sbu,Cavoto:2017otc,Hochberg:2017wce,Knapen:2017ekk,Baryakhtar:2018doz,Szydagis:2018wjp,Griffin:2018bjn,Kurinsky:2019pgb,Acanfora:2019con,Hochberg:2019cyy,Trickle:2019ovy,Caputo:2019cyg,Coskuner:2019odd,Trickle:2019nya,Campbell-Deem:2019hdx,Kozaczuk:2020uzb,Baym:2020uos,Griffin:2020lgd,Trickle:2020oki,Knapen:2020aky,Caputo:2020sys,Campbell-Deem:2022fqm,Liang:2022xbu,vonKrosigk:2022vnf,Berghaus:2022pbu,Krnjaic:2024bdd}, leading to a surge of theoretical activity in sub-GeV DM model building (see, e.g. Refs.~\cite{Knapen:2017xzo,Lin:2019uvt,Zurek:2024qfm}).

An interesting class of sub-GeV DM scenarios accessible by current experimental techniques is based on the absorption of DM in the target material~\cite{Dror:2019onn, Dror:2019dib, Dror:2020czw}. The rest mass of the DM particle is then converted into recoil energy, which becomes as large as $T_r \sim m_\chi^2/2 m_T$, where $m_T$ is the mass of the target and $m_\chi \ll m_T$ the mass of the DM particle. 
Absorption in nuclear targets thus probes DM masses down to ${\cal O}({\rm MeV})$~\cite{Dror:2019dib,PICO:2025rku}, while absorption on electrons in conventional liquid xenon DD experiments such as XENON1T~\cite{XENON:2019gfn}, LZ~\cite{LZ:2019sgr}, XENONnT~\cite{XENON:2020kmp}, PandaX-4T~\cite{PandaX:2018wtu} and DARWIN~\cite{DARWIN:2016hyl} gives sensitivity to DM masses in the sub-MeV range.

The possibility of DM absorption on electrons necessarily implies that DM is unstable. While decays to electrons can be avoided for DM with masses below threshold, $m_\chi \le 2 m_e$, loop-induced decays to photons and neutrinos are unavoidable. These scenarios are constrained by ID through X-ray and $\gamma$-ray telescopes, as well as by their imprints on the cosmological history. Interestingly, DM in this mass range can also be produced in extreme astrophysical environments, such as the hot and dense proto-neutron stars (PNS) formed during core-collapse supernovae (SN).
Although limits on DM couplings have been extensively studied in the literature for light bosonic DM (see e.g. Ref.~\cite{ Chang:2016ntp,Chang:2018rso,Croon:2020lrf,Camalich:2020wac,Caputo:2021rux,Caputo:2022rca,Ferreira:2022xlw, Lella:2022uwi,Caputo:2022rca,Lella:2023bfb, Caputo:2024oqc, Lella:2024dmx, Fiorillo:2025sln}
), similar arguments for dark fermions have been much less explored (for early works see Refs.~\cite{Sutherland:1975dr, Dicus:1976ra,Barbieri:1988av}). In fact, SNe can play a significant role in probing these DM models, opening new search strategies that are complementary to DD, ID, and collider searches~\cite{DeRocco:2019jti,Manzari:2023gkt,Vogl:2024ack, Dev:2025tdv,Choi:2025wbw}.

In this Letter, we present a novel application of the classic SN cooling limit~\cite{Raffelt:1996wa} for sub-MeV fermionic DM coupled to electrons. We focus on the models proposed in Ref.~\cite{Dror:2019onn}, where absorption of DM particles leads to striking signatures in DD experiments.
Following the relevant studies in Ref.~\cite{Dror:2019onn, Dror:2020czw,Ge:2022ius} we parametrize the DM-electron interactions in a model-independent way by the following effective field theory (EFT) Lagrangian 
\begin{align}
\label{eq:EFTLag}
\mathcal L_{\chi}=\sum_X\frac{1}{\Lambda_X^2}\left(\bar e\, \Gamma_X e\right)\cdot(\bar\chi\, \Gamma_X\nu_L) + {\rm h.c.} \, ,
\end{align}
where $e$ and $\nu_L$ are the Standard Model (SM) electron and neutrino fields, $\chi$ is the DM fermion, and the interactions are parametrized by the effective UV scale $\Lambda_X$ and the Dirac matrix $\Gamma_X$.  

\begin{figure*}[t!]
\begin{tabular}{cc}
    \centering    \includegraphics[width=1.0\columnwidth]{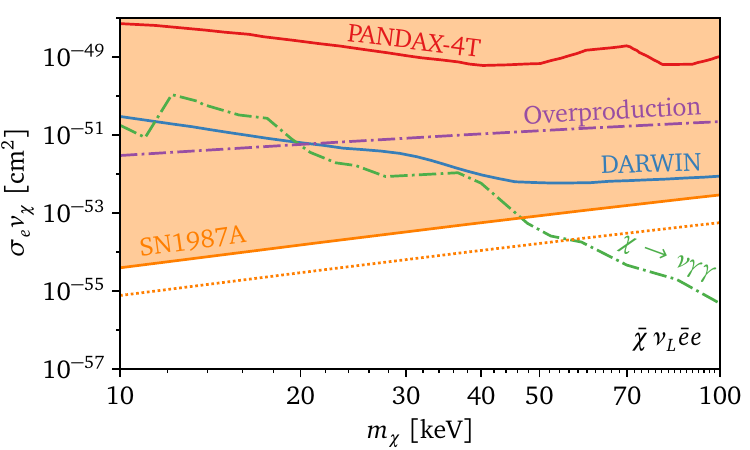} &
\includegraphics[width=1.0\columnwidth]{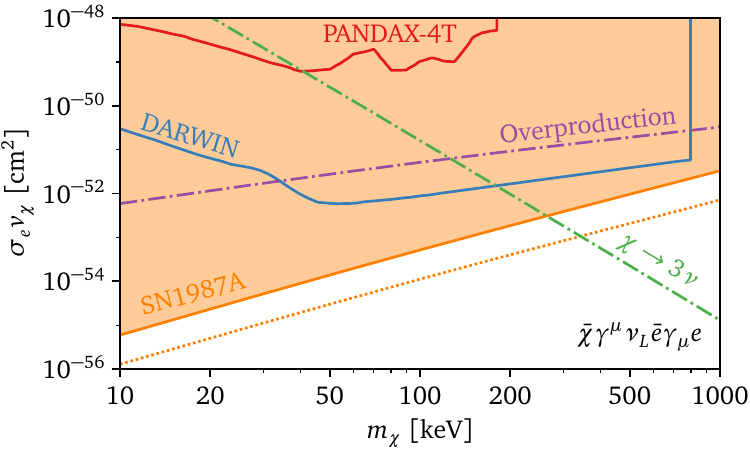}
\end{tabular}    
    \caption{SN~1987A cooling constraints for scalar (left) and vector (right) interactions compared to DD searches by PANDA4T-X and prospects from DARWIN. The dotted 
    line covers the estimated uncertainty from using different SN models. Other upper bounds include cosmological overproduction (see text and SupM for details) and those  from ID derived in Ref.~\cite{Ge:2022ius}. 
    \label{fig:money}}
\end{figure*}

We focus our analysis on the vector operator, $\Gamma_{X}=\Gamma_V=\gamma_\mu$, and scalar operator $\Gamma_{X}=\Gamma_S=\mathbb{1}$, both of which have been probed in DD by PandaX~\cite{PandaX:2022ood,PandaX:2024cic} and CDEX~\cite{CDEX:2024bum}, as well as in ID~\cite{Ge:2022ius}. Our derivation of SN cooling limits can be extended to other operators, including those in which we replace the SM neutrino by a sterile one $\nu_R$~\cite{Dror:2020czw}. In the supplemental material (SupM), we present the analysis of these operators for completeness. 

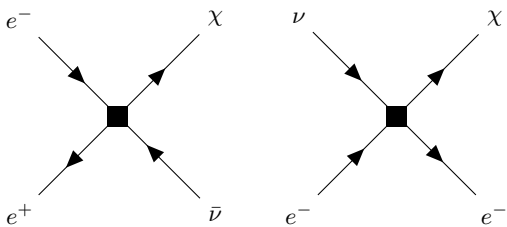
\begin{figure}[tbh!]
    \centering  
    \begin{tikzpicture}
    \begin{feynman}
        \node[shape=rectangle,fill=black] (n) at (0, 0) {\rule{0.05cm}{0.05cm}};
        \vertex (C) at (0,0);
        \vertex (LL) at (-1.3,-1.3) {$e^+$};
        \vertex (HL) at (-1.3,+1.3) {$e^-$};
        \vertex (LR) at (1.3,-1.3) {$\bar{\nu}$};
        \vertex (HR) at (1.3,+1.3) {$\chi$};
        \diagram*{
            (HL) -- [fermion] (C),
            (LL) -- [anti fermion] (C),
            (C) -- [fermion] (HR),
            (C) -- [anti fermion] (LR)
            };
    \end{feynman}
    \end{tikzpicture}
    \begin{tikzpicture}
    \hspace{0.5cm}
    \begin{feynman}
        \node[shape=rectangle,fill=black] (n) at (0, 0) {\rule{0.05cm}{0.05cm}};
        \vertex (C) at (0,0);
        \vertex (LL) at (-1.3,-1.3) {$e^-$};
        \vertex (HL) at (-1.3,+1.3) {$\nu$};
        \vertex (LR) at (1.3,-1.3) {$e^-$};
        \vertex (HR) at (1.3,+1.3) {$\chi$};
        \diagram*{
            (HL) -- [fermion] (C),
            (LL) -- [fermion] (C),
            (C) -- [fermion] (HR),
            (C) -- [fermion] (LR)
            };
    \end{feynman}
    \end{tikzpicture}
\caption{Feynman diagrams for production of DM particles $\chi$ in the PNS for the interactions in Eq.~\eqref{eq:EFTLag}, referred to as \textit{annihilation} (left) and \textit{scattering} (right) processes, respectively.  \label{fig:diagrams}
}
\end{figure}

DM interacting with the SM through 
Eq.~\eqref{eq:EFTLag} can be produced in SNe. A bound on $\Lambda_X$ then follows from the classic SN cooling upper limit: the luminosity induced by the emission of DM particles should be lower than the neutrino luminosity, $L_\chi\lesssim L_\nu$, at 1 sec post-bounce, to be consistent with the neutrino emission signal detected during SN~1987A~\cite{Raffelt:1996wa}. Here, both $L_\nu$ and $L_\chi$ depend on the numerical values of the thermodynamical parameters used in the underlying SN simulation.

The SN~1987A bounds on $\Lambda_X$ can be converted to limits on the scattering cross section probed in DD, as shown in Fig.~\ref{fig:money}. These bounds clearly lie within the region probed by DD experiments and are strongly complementary to other astrophysical ID and cosmological limits. In particular, they exclude the whole range of cross sections accessible to current and future DD searches across the full mass window $10\,\text{keV} \lesssim m_\chi \lesssim 1\,\text{GeV}$. 

For the numerical analysis shown in this figure, we use SN simulations from Ref.~\cite{Bollig:2020xdr}. Specifically, we employ the radial profiles of thermodynamic quantities provided for the coolest and hottest PNS, corresponding to the SFHo18.80 and SFHo20.0 models, respectively. These profiles are used to compute the dark luminosity $L_\chi= \int dV ( Q_{\chi} + Q_{\overline \chi}) $ as a function of the coupling scale $\Lambda_X$, assuming neutrino luminosities of $L_\nu = 5.7 \times 10^{52}~\text{erg/s}$ for SFHo18.80 and $L_\nu = 1.0 \times 10^{53}~\text{erg/s}$ for SFHo20.0. In the following, we describe these calculations and the derivation of our limits in Fig.~\ref{fig:money} in detail.\\

\noindent\textit{Dark luminosity in the free-streaming regime} -- The Feynman diagrams that contribute to the thermal production of $\chi$'s in the core of PNS are shown in Fig.~\ref{fig:diagrams}. If the $\chi$ particles interact weakly, they freely stream out of the inner core, leading to an energy loss rate per unit volume $Q$. Therefore, DM is produced by processes $\psi_1\psi_2\to\psi_3\chi$, where $\psi_i$ are SM fermions and
\begin{align}
Q=& \int \Big[\prod_{i=1}^4 \frac{d^3\vec p_i}{(2\pi)^3 2E_i}\Big] 
(2\pi)^4 \delta^4 (p_1+p_2-p_3-p_\chi) \nonumber \\
& \times f_1 f_2(1-f_3)\sum_{\rm spins}|\mathcal{M}|^2 E_\chi \,.
\label{eq:FSgeneral}
\end{align}
The $p_i=(E_i,\vec p_i)$ are the 4-momentum of the particles labeled by $i$, with $i=4$ corresponding to the $\chi$.
The squared matrix elements $|\mathcal{M}|^2$, summed over all fermion spins, encode the interactions that produce the scattering process, while the Fermi-Dirac occupation numbers $f_i^{-1} = \exp{(E_i-\mu_i})/T + 1$ describe the local themodynamic equilibrium conditions of the PNS parametrized by the chemical potential $\mu_i$ and temperature $T$.

The energies $E_i$ are defined in the rest frame of the PNS, where the initial particles collide with 3-momenta $\vec p_1$ and $\vec p_2$ forming an angle $\theta$. The energy loss rate can then be factorized as (cf.~Ref.~\cite{Manzari:2023gkt})
\begin{align}
\label{eq:Q}
Q& =\frac{1}{32\pi^4}\int^{\infty}_{m_1}dE_1 \bar p_1 f_1\int^{\infty}_{m_2}dE_2 \bar p_2f_2\int^{1}_{-1}dc_\theta \Theta_s J_s \, , 
\end{align}
where $\bar p_i=|\vec p_i|$, $m_i$ are the masses and $c_\theta=\cos\theta$. The function $J_s$ contains the dynamical information on the scattering process  
\begin{align}
\label{eq:Js1}
J_s (E_1, E_2, s) & =\frac{ \bar p^{\, \prime}_{34}}{16\pi^2\sqrt{s}}\int ^1_{-1}dc_{\theta^\prime}\int^{2\pi}_0d\phi^\prime (1-f_3) \nonumber \\
& \times \sum_{\rm spins} |\mathcal M (s,t)|^2 E_\chi \, .
\end{align}
where $s = (p_1 + p_2)^2$ and $t = (p_1 - p_3)^2$ are the Mandelstam variables and the Heavyside function $\Theta_s \equiv \theta (s-(m_1+m_2)^2)\theta(s-(m_3+m_\chi)^2)$ enforces kinematic thresholds. In this equation, primed variables refer to the center-of-mass frame and $\bar p^{\prime}_{34}$ is the corresponding absolute value of the final state 3-momentum. The variables in the PNS frame $E_3$ (hidden in $f_3$) and $E_\chi$ can be expressed as functions of $E_1$, $E_2$, $s$ before integrating over the angles $\theta^\prime$ and $\phi^\prime$, see SupM for more details. 

Significant simplifications of Eq.~\eqref{eq:Js1} can be obtained by employing two approximations: 
\begin{enumerate}
\item \textbf{Degeneracy factorization:} replacing the Pauli-blocking factor $(1-f_i)$ by its thermal average~\cite{Raffelt:1996wa}
\begin{align}
(1-f_i) \to  F_i\equiv \frac{\mathfrak g_i}{n_i}\int \frac{d^3 \vec {p}_i}{(2\pi)^3}f_i(1-f_i) \, .
\end{align}
where $\mathfrak g_i$ is the number of degrees of freedom and $n_i$ the number density. For \textit{typical} PNS conditions ($T=30$ MeV, $\mu_e=130$ MeV,  $\mu_{\nu_e}=20$ MeV) $F_{e^-} = 0.53, F_{\nu_e} = 0.86$, which is a moderate effect.
\item \textbf{Massless limit:} all particles can be taken as massless, since for the DM mass range of interest $m_\nu, m_e, m_\chi \ll T,\;\mu_i$ in SN.  
\end{enumerate} 

These approximations lead to simple analytical formulae for the contributions to $Q$ from the two processes in Fig.~\eqref{fig:diagrams}. For the \textit{annihilation} process $e^+e^-\to\nu\bar\chi$ ($e^+e^-\to\bar\nu \chi$) one finds
\begin{align}
\label{eq:Q0_ann}
Q_{\ann}=\frac{a_X\,T^9  F_{\nu (\overline \nu)}}{18\pi^5\Lambda_X^4}  H_4(y_e)H_3(-y_e)+ \left( y_e \to - y_e\right) \, , 
\end{align}
where $y_i=\mu_i/T$, $a_V=1$ or $a_S=3/4$, and 
\begin{align}
\label{eq:Hn}
H_n(y)=\int^\infty_0dx \frac{x^n}{1+e^{x-y}}=-n!\, {\rm Li}_{n+1}\left(-e^y\right) \, .
\end{align}
For the \textit{scattering} process $e^-\nu\to e^-\chi$ one obtains
\begin{align}
\label{eq:Q0_scatt}
Q_{\scat}&=\frac{T^9 F_{e^-}}{72\pi^5\Lambda_{X}^4} \left[ b_XH_4(y_e)H_3(y_\nu)+c_XH_4(y_\nu)H_3(y_e)\right] \,,
\end{align}
where $b_V=7$, $c_V=9$, and $b_S=3/2$, $c_S=1/2$. 
Other channels like $e^-\bar\nu\to e^-\bar \chi$ or $e^+\nu\to e^+ \chi$ are obtained from Eq.~\eqref{eq:Q0_scatt} by replacing the arguments of the $H_n(y)$ functions accordingly. 
Compared to annihilation in Eq.~\eqref{eq:Q0_ann}, the contribution of electron scattering to the energy-loss rate does not suffer the strong suppression due to the small positron abundance, and is only slightly suppressed by partial electron degeneracy. Indeed, for \textit{typical} PNS conditions
$H_4 (y_e) \approx 1040$, $H_3 (y_e) \approx 192$ but $H_4 (-y_e) \approx 0.3$, $H_3 (-y_e) \approx 0.08$, compared to the neutrino conditions $H_4 (y_{\nu_e}) \approx 44$ and $H_3 (y_{\nu_e}) \approx 11$. Therefore, neutrino scattering on electrons provides the leading contribution to $Q$, while scattering on anti-neutrinos gives a rate smaller by about a factor 4, for nominal conditions in the PNS. 

Adding the contributions in $Q_{\ann}$ and $Q_{\scat}$ one obtains
\begin{align}
\label{eq:Lambda_FS}
&\Lambda_S  \gtrsim  \left(9.3 - 14 \right)\TeV \, , \nonumber\\
&\Lambda_V  \gtrsim \left(15  - 22\right)  \TeV \, , 
\end{align}
where the lower (upper) value correspond to the colder (hotter) simulation SFHo18.80 (SFHo20.0)~\cite{Bollig:2020xdr}.\\

\noindent\textit{Dark luminosity in the trapping regime} -- 
In the trapping regime the DM particles reach thermal equilibrium with the plasma in the PNS and are emitted from a surface with radius $r_\chi$ (the \textit{dark sphere}) following a law analogous to black-body radiation. Including the degrees of freedom of approximately massless $\chi$ and $\bar\chi$, this reads
\begin{align}
L_{\chi}^{\rm trap} = \frac{7\pi^3}{30} r_\chi^2 T_\chi^4 \,  ,
\label{eq:blackbody}
\end{align}
where
$T_\chi = T( r_\chi)$ and the radius $r_\chi$ is defined by requiring the optical depth to be $\tau_\chi (r_\chi) = \int_{r_\chi}^\infty dr/\lambda(r)=2/3$~\cite{UBHD-66179479,UBHD-35002838,UBHD-68454320},
where $\lambda(r)$ is a suitable spectral average of the DM's mean free path (MFP) at a radius $r$. Here we use a ``naive'' thermal average 
\begin{align}
\label{eq:trapp}
\lambda (r) & = 
 \frac{\mathfrak g_\chi}{n_\chi} \int \frac{d^3\vec p_\chi}{(2\pi)^3} f_\chi\lambda(r,E_\chi) \, . 
\end{align}
The energy-dependent MFP $\lambda (r, E_\chi)$ is related to the total absorption rate of DM particles in the medium, $\Gamma=1/\lambda$. Given the thermodynamical conditions of the plasma at radius $r$, the rate for $\chi\psi_1 \to \psi_2\psi_3$  is
\begin{align}
\Gamma(E_\chi)=& \frac{1}{2 E_\chi \mathfrak  g_\chi}\int \Big[\prod_{i=1}^3 \frac{d^3\vec p_i}{(2\pi)^3 2E_i}\Big]  f_1 (1-f_2)(1-f_3)
 \nonumber \\
& \times (2\pi)^4 \delta^4 (p_1+p_2-p_3-p_\chi)\sum_{\rm spins}|\mathcal{M}|^2 \,,
\label{eq:WidthGeneral}
\end{align}
where the particle $i=1$ is identified with the SM scatterer and $i=2,3$ with the final SM particles, and where $\mathcal M$ is the scattering amplitude of the process.
Following an analysis similar to Eq.~\eqref{eq:Q} we find
\begin{align}
\label{eq:Gamma0}
\Gamma(E_\chi)& =\frac{1}{16\pi^2E_\chi\mathfrak g_\chi}\int^{\infty}_{m_1}dE_1 \bar p_1 f_1\int^{1}_{-1}dc_\theta \Theta_s K_s \, , 
\end{align}
where $\Theta_s \equiv \theta (s-(m_2+m_3)^2)\theta(s-(m_1+m_\chi)^2)$ and 
\begin{align}
\label{eq:Ks1}
K_s (E_\chi, E_1, s) & =\frac{ \bar p^{\, \prime}_{23}}{16\pi^2\sqrt{s}}\int ^1_{-1}dc_{\theta^\prime}\int^{2\pi}_0d\phi^\prime (1-f_2)(1-f_3) \nonumber \\
& \times \sum_{\rm spins} |\mathcal M (s,t)|^2  \, .
\end{align}
Using the same approximations as for $Q$ one obtains
\begin{align}
&\Gamma_{\ann}(E_\chi)=\frac{2a_XE_\chi T^4F_{e}F_{\bar e}} {9\pi^3 \Lambda_X^4 \mathfrak g_\chi}H_3(-y_\nu),\nonumber\\
&\Gamma_{\scat}(E_\chi)=\frac{4d_X E_\chi T^4F_{\nu}F_{e}}{9\pi^3\Lambda_X^4\mathfrak g_\chi}H_3(y_e),
\end{align}
for \textit{inverse annihilation}, $\chi \bar\nu\to e^+ e^-$, and \textit{inverse scattering}, $\chi e^-\to \nu e^-$ (with coeffcients $d_V=1$ and $d_S=1/8$), respectively and where we have assumed that DM is in chemical equilibrium,  $\mu_\chi=\mu_\nu$.
One finds that the SN~1987A constraint has a lower bound at
\begin{align}
\label{eq:Lambda_TR}
&\Lambda_S  \lesssim \left( 38 - 41 \right) \GeV \, , \nonumber\\
&\Lambda_V  \lesssim \left( 56 - 59  \right) \GeV \, , 
\end{align}
for the simulations SFHo18.80 and SFHo20.0~\cite{Bollig:2020xdr}.\\

\noindent\textit{Results and discussion} -- The SN~1987A cooling limits derived above can be converted to upper limits on the DD cross section predicted in the EFT~\cite{Ge:2022ius}. The results for vector and scalar operators shown in Fig.~\ref{fig:money} demonstrate that the new SN bound covers a large fraction of the previously unconstrained region of interest for ID, and current and future DD experiments (i.e. DARWIN~\cite{DARWIN:2016hyl}). 

The cosmological overabundance limits in Fig.~\ref{fig:money} are obtained assuming that DM is produced in the early Universe via UV freeze-in by the \emph{same} mechanisms as in SN, \textit{cf.} Fig.~\ref{fig:diagrams}. Our limits, derived in the SupM, are more stringent than those reported in previous studies~\cite{Dror:2020czw,Ge:2022ius}. 

\begin{figure}[t!]
    \centering    \includegraphics[width=1.\linewidth]{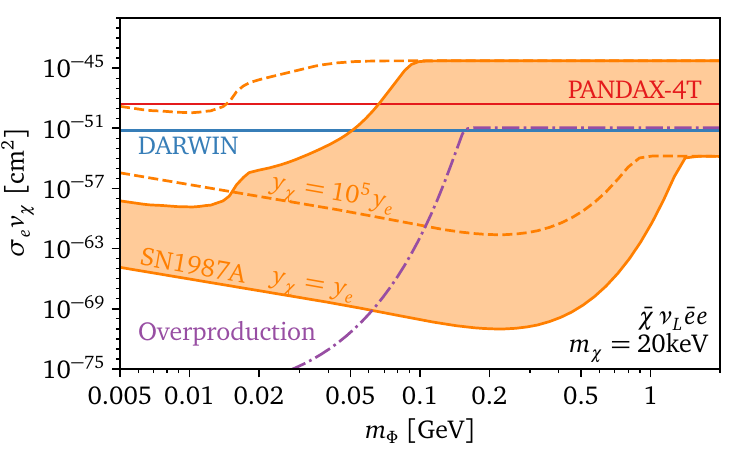}
    \caption{Bounds on the DD cross-section for the absorption of DM with $m_\chi=20$ keV and sub-GeV scalar mediator. For $y_e=y_\chi$,  the SN~1987A limits from SFHo18.80 are compared to the  limit from DM overproduction 
    and sensitivity limits achieved by PANDAX-4T and the prospects for DARWIN. We also show the SN~1987A bound obtained using $y_\chi=10^5 y_e$ or vice-versa. For such hierarchical couplings the overproduction bound would only get stronger.
}
    \label{fig:light_scalar}
\end{figure}

One important caveat in our analysis, concerning the validity of the results shown in Fig.~\ref{fig:money}, is the use of EFT. While this approximation holds in DD and ID, where the relevant energy scale (set by the DM mass) is assumed to be much smaller than the mediator mass, it can break down in the context of thermal production of DM in the PNS, where temperatures can reach $\mathcal{O}(10\,\text{MeV})$.

In Fig.~\ref{fig:light_scalar} we show the limits on a simplified UV completion of the scalar operator, with interactions mediated by a real scalar $\Phi$,
\begin{align}
\label{eq:Lag_scalar_med}
\mathcal L_S\supset y_e \bar e e\Phi+y_\chi\bar\chi\nu_L\Phi+\text{h.c.} \, ,
\end{align}
for a given DM mass $m_\chi=20$ keV and as a function of the mediator mass $m_\Phi$. For heavy mediators above $\approx 1$ GeV ($\approx100$ MeV) the SN cooling limits become independent of the mediator mass and approach the results of the EFT calculation for the free-streaming (trapping regime). However, for light mediators the bound departs strongly from the EFT behavior. The physical reason for this change is the onset of resonant production processes in annihilation $e^+ e^-\to\Phi^*\to\chi\bar\nu$ and photo-production $e^-\gamma \to e^-\Phi^*(\to\chi\bar\nu)$~\cite{Manzari:2023gkt}, described for completeness in the SupM. The production and absorption rates in SN in this resonant regime scale quadratically (and not quartically) with the couplings
and the bound covers a region of couplings much smaller than what could have been naively expected from the EFT. Therefore, as shown in Fig.~\ref{fig:light_scalar}, SN~1987A does not constrain the region accessible to DD for light mediators with $m_\Phi\lesssim70$ MeV, unless there are large hierarchies in the couplings. 

However, as stressed above, the same processes underlying DM production in the PNS will produce DM in the early Universe through freeze-in. The temperatures dominating the cosmological DM production are around the reheating temperature, $T_{R}\gtrsim5$ MeV,
which are similar to those in SN, and cosmological DM production is also resonantly enhanced for light mediators~\cite{Frangipane:2021rtf}. This is shown in Fig.~\ref{fig:light_scalar} where the overproduction bound is drastically strengthened due to this resonant behavior for $m_\Phi\lesssim150$ MeV, excluding the region of interest for DD. 

In conclusion, in this work, we have presented new SN~1987A cooling bounds on sub-MeV fermionic dark matter with effective couplings to electrons. Importantly, our bounds exclude a significant fraction of the projected sensitivity regions of both current and future DD experiments and are highly complementary to ID searches with X-ray and $\gamma$-ray telescopes and cosmological limits. We have 
corroborated these conclusions obtained in the EFT by exploring how these constraints are affected in simple UV completions with sub-GeV mediators. While these mediators can be produced on-shell in supernova cores, shifting the constrained regions to smaller couplings, the same resonant behavior occurs in the early Universe, strengthening the overproduction limits. As a result, the parameter space accessible to direct detection remains excluded.

Our results highlight the important role that SN analyses can play in probing sub-GeV DM scenarios beyond the reach of traditional laboratory and cosmological searches.

\let\oldaddcontentsline\addcontentsline
\renewcommand{\addcontentsline}[3]{}

\section*{Acknowledgments}
We thank Filippo Sala for useful discussions and Robert Bollig and Hans Thomas Janka for sharing data of the simulations with us. JMC also acknowledges CERN-TH for hospitality during the completion of this article. This work is partially supported by project B3a of the DFG-funded Collaborative Research Center TRR257, ``Particle Physics Phenomenology after the Higgs Discovery" and has received support from the European Union's Horizon 2020 research and innovation programme under the Marie Sk\l{}odowska -Curie grant agreement No 860881-HIDDeN. The work of C.A.M. is supported by the Office of High Energy Physics of the U.S. Department of Energy under contract DE- AC02-05CH11231. JMC acknowledge support from the MICINN through the grant ``DarkMaps'' PID2022-142142NB-I00 and from the European Union through the grant ``UNDARK'' of the Widening participation and spreading excellence programme (project number 101159929). J.S.\ is funded by the Carl-Zeiss-Stiftung through the project \textsl{Model-Based AI: Physical Models and Deep Learning for Imaging and Cancer Treatment}.

\textit{Note added --} While this work was being finalized, an independent analysis of SN~1987A bounds on the same effective models appeared as a preprint~\cite{Lin:2025mez}. Compared to their results, our bounds are approximately one to two orders of magnitude weaker (depending on the SN simulation used).
Our supernova analysis is more systematic, extends to the  case of simple UV completions and sterile neutrinos, and we derive consistently the relevant cosmological limits from overproduction. In particular, we discuss the implications of light mediators where the EFT analysis breaks down.

\bibliography{bib}

\let\addcontentsline\oldaddcontentsline

\clearpage

\onecolumngrid
\begin{center}
  \textbf{\large Supplementary Material for The SN1987A Cooling Bound on Dark Matter Absorption in Electron Targets
  }\\[.2cm]
  \vspace{0.05in}
  {Claudio Andrea Manzari, Jorge Martin Camalich, Jonas Spinner, and Robert Ziegler}
\end{center}

\twocolumngrid

\setcounter{equation}{0}
\setcounter{figure}{0}
\setcounter{table}{0}
\setcounter{section}{0}
\setcounter{page}{1}
\makeatletter
\renewcommand{\theequation}{S\arabic{equation}}
\renewcommand{\thefigure}{S\arabic{figure}}
\renewcommand{\thetable}{S\arabic{table}}

\setcounter{secnumdepth}{2}
\renewcommand{\thesection}{\Roman{section}}
\renewcommand{\thesubsection}{\thesection.\alph{subsection}}

\onecolumngrid

\startcontents[sections]
\tableofcontents

\section{Kinematics in CM and PNS Frames}\label{app:kinematics}

\noindent Here we discuss the setup for the production of DM in the PNS medium through the annihilation and scattering processes in Fig.~\ref{fig:diagrams}. We choose a coordinate system where the 4-momenta of the incoming $\psi_1$ and $\psi_2$ particles in the PNS frame read 
\begin{align}
p_1& =(E_1,0,0,\bar p_1) \, , \ & p_2 & =(E_2,\bar p_2 s_\theta,0,\bar p_2 c_\theta) \, , 
\end{align}
with $\bar p_i \equiv |\vec p_i|=\sqrt{E_i^2-m_i^2}$ and $s_\theta = \sin \theta, c_\theta = \cos \theta$. In the center-of-mass (CM) frame the incoming $\psi_1$ and $\psi_2$ particles collide along the $z$-axis with 4-momenta
\begin{align}
p_1^\prime & =(E_1^\prime,0,0,\bar p^\prime_{12}) \, , & p_2^\prime & =(E_2^\prime,0,0,-\bar p^{\prime}_{12}) \, , 
\end{align}
where
\begin{align}
\bar p^{\prime}_{12}  & =\frac{1}{2\sqrt{s}}\lambda^{1/2}(s,m_1^2,m_2^2) \, ,  & 
E_1^\prime & =\frac{1}{2\sqrt{s}}\left(s+m_1^2-m_2^2\right) \, ,  &E_2^\prime & =\frac{1}{2\sqrt{s}}\left(s-m_1^2+m_2^2\right) \, , 
\end{align}
with $\lambda(a,b,c)=a^2+b^2+c^2-2(ab+ac+bc)$. The outgoing particles $\psi_3$ and $\chi$ scatter under a polar angle $\theta^\prime$ with respect to the $z$-axis and within a plane $\mathcal P^\prime$ that forms an azimuthal angle $\phi^\prime$ with respect to the $x-z$ plane: 
\begin{align}
p_3^\prime & =(E_3^\prime,\bar p^{\prime}_{34} \left(s_{\theta^\prime} c_{\phi^\prime}, s_{\theta^\prime} s_{\phi^\prime}, c_{\theta^\prime}) \right) \, , & p_4^\prime &=(E_4^\prime,-\bar p^{\prime}_{34} \left(s_{\theta^\prime} c_{\phi^\prime}, s_{\theta^\prime} s_{\phi^\prime}, c_{\theta^\prime}) \right) \, , 
\end{align}
where
\begin{align}
\bar p^{\prime}_{34} & =\frac{1}{2\sqrt{s}}\lambda^{1/2}(s,m_3^2,m_\chi^2)\, , & E_3^\prime & =\frac{1}{2\sqrt{s}}\left(s+m_3^2-m_\chi^2\right) \, ,  &E_4^\prime & =\frac{1}{2\sqrt{s}}\left(s-m_3^2+m_\chi^2\right) \, .
\end{align}
The Mandelstam variable $t$ evaluated in in the CM frame can be written in terms of $s$ and the polar angle $\theta^\prime$,
\begin{align}
\label{eq:t_CM}
t  =  (p_2 - p_4)^2 =  -\frac{1}{2s}\left(s^2+m_1^2m_3^2 +m_2^2m_\chi^2 - m_1^2m_\chi^2 - m_2^2 m_3^2 \right)+ \frac{1}{2} \left(m_1^2+m_2^2+m_3^2+m_\chi^2\right) +  2 \,\bar p^{\prime}_{12} \bar p^{\prime}_{34}c_{\theta^\prime} \, .
\end{align}
The Lorentz transformation from the PNS frame to the CM frame is given by 
\begin{align}
\label{eq:Lorentz_tranfs}
&\vec \beta=\frac{\bar p}{E}(s_\eta,0,c_\eta) \, , & &\gamma=\frac{E}{\sqrt{s}} \, , &
\end{align}
where we have defined
\begin{align}
&E=E_1+E_2 \, , & \bar p=|\vec p_1+\vec p_2|=\sqrt{E^2-s} \, , 
\end{align}
with  the angle $\eta$  given by
\begin{align}
c_\eta=\frac{\sqrt{s}E_1-E E_1^\prime}{\bar p \bar p^{\prime}_{12}} \, .
\end{align}
This relation allows the final state energies in the PNS frame to be expressed in terms of the kinematic variables as
\begin{align}
\label{eq:PNSinCM}
E_3 & =\frac{1}{\sqrt{s}}\left(E E_3^\prime+ \bar p \bar p^{\prime}_{34}\left( s_{\theta^\prime} c_{\phi^\prime} s_\eta+c_{\theta^\prime} c_\eta\right)\right) \, , &
E_4 & =\frac{1}{\sqrt{s}}\left(E E_4^\prime- \bar p \bar p^{\prime}_{34}\left( s_{\theta^\prime} c_{\phi^\prime} s_\eta+c_{\theta^\prime} c_\eta\right)\right) \, .
\end{align}
With these definitions, one obtains Eqs.~\eqref{eq:Js1} and \eqref{eq:Ks1}.

\section{Scattering Amplitudes, Cross-sections, Cooling Rates }
Here we provide results for squared matrix elements, cross-sections and the approximate cooling rates for general effective operators defined by (note that only one tensor operator is independent)    
\begin{align}
\label{eq:EFT_Lag}
\mathcal L_{\chi}&=\frac{1}{\Lambda_V^2}(\bar e\gamma^\mu e)(\bar \chi\gamma_\mu\nu_L)+\frac{1}{\Lambda_A^2}(\bar e\gamma^\mu\gamma_5 e)(\bar \chi\gamma_\mu\nu_L)+ 
\frac{1}{\Lambda_S^2}(\bar e e)(\bar\chi \nu_L)+\frac{1}{\Lambda_P^2}(\bar e\gamma_5 e)(\bar\chi \nu_L)+\frac{1}{\Lambda_T^2}(\bar e\sigma^{\mu\nu} e)(\bar\chi \sigma_{\mu\nu}\nu_L)+{\rm h.c.} 
\end{align}
We will first discuss annihilation processes and then scattering, turning on a single operator in Eq.~\eqref{eq:EFT_Lag} at a time. 
\subsection{Annihilation}
The spin-summed (but not spin-averaged) squared matrix elements for annihilation processes $e^+e^-\to\chi \overline \nu$ and $e^+e^-\to\overline \chi \nu$ are given by
\begin{align}
\label{eq:M2_anni}
\sum_{\rm spins} |\mathcal M_{\ann}^V|^2= & \frac{4}{\Lambda_V^4} \left(2m_e^4-4 m_e^2 t+2t \left(s+t-m_\chi^2\right) +  s \left(s-m_\chi^2\right) \right) \, , \\
\sum_{\rm spins} |\mathcal M_{\ann}^A|^2 = & \frac{4}{\Lambda_A^4} \left(2m_e^4+4 m_e^2(m_{\chi}^2-s-t) -m_\chi^2(s+2t)  + s^2 +2\,s\,t +2\,t^2 \right) \, ,\\
\sum_{\rm spins} |\mathcal M_{\ann}^S|^2 = & \frac{2}{\Lambda_S^4} (s-4 m_e^2)(s - m_\chi^2) \, , \\
\sum_{\rm spins} |\mathcal M_{\ann}^P|^2 = &\frac{2}{\Lambda_P^4}s(s - m_\chi^2) \, , \\
\sum_{\rm spins} |\mathcal M_{\ann}^T|^2 = &\frac{16}{\Lambda_T^4}\left(4 m_e^2 + 2 m_e^2(m_{\chi}^2-s-4t)-m_{\chi}^2(s+4t) + (s+2t)^2\right) \, , 
\end{align}
with the Mandelstam variables
\begin{align}
s & = 2 m_e^2 + 2 (E_1 E_2 - \bar p_1 \bar p_2 c_\theta) \, , & 
 t & = (p_{e^-} - p_{\chi (\overline \chi)})^2 = m_e^2-\frac{s-m_\chi^2}{2}\left(1-\beta_e\,c_{\theta^\prime}\right) \, , & \beta_e & =\sqrt{1-\frac{4m_e^2}{s}} \, ,
\end{align}
and $\theta^\prime$ ($\theta$) are the scattering angles in the CM (PNS) frame, see SupM~\ref{app:kinematics}. The corresponding cross-sections read 
\begin{align}
\sigma_{e^+ e^- \to \chi \overline \nu}^V & =  \frac{\beta_e }{48\pi\Lambda_V^4s^2(s-4m_e^2)} (s+2m_e^2)(s-m_\chi^2)^2(2s+m_\chi^2)\,,\\
\sigma_{e^+ e^- \to \chi \overline \nu}^A & =\frac{\beta_e }{48\pi\Lambda_A^4 s^2(s-4m_e^2)} (s-m_\chi^2)^2 \left(s(2s+m_\chi^2) + 2m_e^2(m_\chi^2-4s)\right)\,,\\
\sigma_{e^+ e^- \to \chi \overline \nu}^S & = \frac{\beta_e }{32\pi\Lambda_S^4s}(s-m_\chi^2)^2\,,\\
\sigma_{e^+ e^- \to \chi \overline \nu}^P & = \frac{\beta_e }{32\pi\Lambda_P^4 (s-4m_e^2)}(s-m_\chi^2)^2\,,\\
\sigma_{e^+ e^- \to \chi \overline \nu}^T & = \frac{\beta_e }{12 \pi\Lambda_T^4s^2(s-4m_e^2)}(s+2m_e^2)(s-m_\chi^2)^2(s+2m_\chi^2)\,,
\end{align}
and the inverse processes can be obtained using the relation 
\begin{align}
\sigma_{\chi \overline \nu \to e^+ e^-}^X =  \frac{4 s(s-4m_e^2)}{\mathfrak g_\chi (s-m_\chi^2)^2}\sigma_{e^+ e^- \to \chi \overline \nu}^X \, ,
\end{align}
for $X = V,A,P,S,T$ and $g_\chi$ denotes the number of degrees of freedom in $\chi$. 
Employing the approximation $(1-f_3)\to F_{\overline \nu}$ (for  $e^+e^-\to\chi \overline \nu$), one can easily calculate $J_{s,\rm ann}^X$ using Eq.~\eqref{eq:Js1} as the integrals  $\phi^\prime$ and $c_\theta^\prime$ are trivial.
Taking the massless limit, $m_e= m_\chi=0$, gives 
\begin{align}
J^{X}_{s,{\ann}}= s^2 \frac{E_1+E_2}{6\pi\Lambda_X^4}a_X F_{\overline \nu}  \, , 
\label{Jhatsann}
\end{align}
where $a_V=a_A=4a_S/3=4a_P/3=a_T/2=1$. The analogous expressions for $e^+e^-\to \overline \chi \nu$ are identical except for the Pauli blocking factor $(1-f_3)$ that instead contains the chemical potential of the neutrino and becomes $F_{\nu}$ with our approximations. This leads to 
\begin{align}
Q_{\ann}^X=\frac{a_X T^9  F_{\nu (\overline \nu)}}{18\pi^5\Lambda_X^4}  H_4(y_e)H_3(-y_e)+ \left( y_e \to - y_e\right) \, , 
\end{align}
where $y_i=\mu_i/T$ and using Eq.~\eqref{eq:Hn}. 
Likewise, the absorption width and rates for the process $\chi\bar\nu \to e^+ e^-$, in the $m_e = m_{\chi} = 0$ limit, are
\begin{align}
K^{X}_{\ann} =& \frac{s^2}{3\pi\Lambda_X^4}a_XF_eF_{\bar{e}}  \, ,\\
\Gamma^{X}_{\ann} = &\frac{2E_{\chi}T^4H_3(-y_\nu)}{9\, \mathfrak{g}_{\chi}\pi^3\Lambda_X^4}a_XF_eF_{\bar{e}}  \, ,\\
\mathcal C_{\ann}^X=& \frac{T^8H_3(-y_\nu)H_3(y_\nu)}{9\pi^5\Lambda_X^4}a_XF_eF_{\bar{e}}  \, ,
   \label{Cann}
\end{align}
where we have defined the collision operator,
\begin{align}
\mathcal C=\mathfrak g_\chi\int \frac{d^3\vec p_\chi}{(2\pi)^3}f_\chi \Gamma(E_\chi).
\end{align}

\subsection{Scattering}
The spin-summed (but not spin-averaged) squared matrix elements for  scattering processes $e^\mp\nu\to e^\mp\chi$ or $e^\mp\bar\nu\to e^\mp\bar\chi$ are given by
\begin{align}
\sum |\mathcal M_{\scat}^V|^2= &\frac{4}{\Lambda_V^4} \left(2 s^2+2 s t+t^2-2 m_e^2 \left(2 s-  m_e^2\right)  -m_\chi^2 (2 s+t)\right) \, , \nonumber\\
\sum |\mathcal M_{\scat}^A|^2=&\frac{4}{\Lambda_A^4} \left(4 m_e^2 \left(m_{\chi }^2-s-t\right)+2 m_e^4  -m_{\chi }^2 (2 s+t)+2 s^2+2 s t+t^2\right) \, , \nonumber\\
\sum |\mathcal M_{\scat}^S|^2= &\frac{2}{\Lambda_S^4}\left(t-4m_e^2\right)\left(t-m_\chi^2\right) \, ,  \nonumber\\
\sum |\mathcal M_{\scat}^P|^2 = & \frac{2}{\Lambda_P^4}t\left(t-m_\chi^2\right) \, ,  \nonumber\\
\sum |\mathcal M_{\scat}^T|^2= &\frac{16}{\Lambda_T^4} \left(2 m_e^2 \left(m_{\chi }^2-4 s-t\right)+4 m_e^4  -m_{\chi }^2 (4 s+t)+(2 s+t)^2\right) \, ,
\end{align}
with the Mandelstam variable 
\begin{align}
 t & =  \frac{m_\chi^2}{2} \left(1+ \frac{m_e^2}{s} \right) - \frac{s - m_e^2}{2} \left( 1- \frac{m_e^2}{s} - \beta_{e \chi} c_{\theta^\prime} \right) \, , & \beta_{e\chi} & =   \sqrt{1 -2 \frac{m_e^2 + m_\chi^2}{s} + \frac{(m_e^2 - m_\chi^2)^2}{s^2}} \, ,
\end{align}
and $\theta^\prime$ is the scattering angle in the CM frame, see SupM~\ref{app:kinematics}. The corresponding cross-sections read 
\begin{align}
\begin{split}
\sigma_{e^\mp \nu\to e^\mp \chi}^V & =  \frac{\beta_{e\chi}}{48\pi\Lambda_V^4s^2} \Big(8s^3-s^2(12m_e^2+7m_\chi^2) -s(m_\chi^4-6m_e^4+3m_e^2m_\chi^2)-2m_e^2(m_e^2-m_\chi^2)^2\Big)\,,\\
\sigma_{e^\mp \nu\to e^\mp \chi}^A & = \frac{\beta_{e\chi}}{48\pi\Lambda_A^4s^2} \Big(8s^3-7m_\chi^2s^2-s(m_\chi^4+6m_e^4-9m_e^2m_\chi^2) -2m_e^2(m_e^2-m_\chi^2)^2\Big)\,,\\
\sigma_{e^\mp \nu\to e^\mp \chi}^S & = \frac{\beta_{e\chi}}{96\pi\Lambda_S^4s^2} \Big(2s^3-s^2(m_\chi^2-6m_e^2)-s(6m_e^4+m_\chi^4-9m_e^2m_\chi^2)-2m_e^2(m_e^2-m_\chi^2)^2\Big)\,,\\
\sigma_{e^\mp \nu\to e^\mp \chi}^P & =\frac{\beta_{e\chi}}{96\pi\Lambda_P^4s^2} \Big(2s^3-s^2(6m_e^2+m_\chi^2)-s(m_\chi^4-6m_e^4+3m_e^2m_\chi^2) -2 m_e^2(m_e^2-m_\chi^2)^2\Big)\,,\\
\sigma_{e^\mp \nu\to e^\mp \chi}^T & = \frac{\beta_{e\chi}}{12\pi\Lambda_T^4s^2} \Big(14s^3-s^2(12m_e^2+13m_\chi^2)-sm_\chi^2(m_\chi^2-3m_e^2)-2m_e^2(m_e^2-m_\chi^2)^2\Big)\,,\\
\end{split}
\end{align}
and the inverse processes can be obtained using the relation 
\begin{align}
\sigma_{e^\mp \chi\to e^\mp \nu}^X = \frac{1}{\mathfrak g_\chi} \frac{(s-m_e^2)^2}{s^2-2s(m_e^2+m_\chi^2) + (m_e^2-m_\chi^2)^2}\sigma_{e^\mp \nu\to e^\mp \chi}^X \, .
\end{align}
Using the approximation $(1-f_3)\to F_{e^\pm}$ ($F_{e^-}$ for  $e^- \nu\to e^- \chi$ and $e^- \bar\nu\to e^- \bar\chi$ and $F_{e^+}$ for  $e^+ \nu\to e^+ \chi$ and $e^+ \bar\nu\to e^+ \bar\chi$)
), with $E_1 = E_{e^\pm}, E_2 = E_{\nu/\overline \nu}$,
and taking the massless limits, $m_e= m_\chi=0$, we obtain 
\begin{align}
J_{s,{\scat}}^{V,A} & =\frac{s^2}{24\pi\Lambda_{V,A}^4}(7E_1 + 9E_2) F_{e^\pm}\, , \\
\hat J_{s,{\scat}}^{S,P} & =\frac{s^2}{48\pi\Lambda_{S,P}^4}(3E_1 + E_2) F_{e^\pm}\, , \\
J_{s,{\scat}}^{T} & =\frac{s^2}{6\pi\Lambda_{T}^4}(11E_1 + 17E_2)F_{e^\pm} \, ,
\end{align}
which permits to analytically solve the final integrals to calculate the energy loss rates in this limit for  $e^-\nu\to e^- \chi$:
\begin{align}
Q_{\scat}^{V,A}&=\frac{T^9 F_{e^-}}{72\pi^5\Lambda_{V,A}^4} \left[ 7H_4(y_e)H_3(y_\nu)+9H_4(y_\nu)H_3(y_e)\right] \, , \nonumber\\
Q_{\scat}^{S,P}&=\frac{T^9 F_{e^-}}{144\pi^5\Lambda_{S,P}^4} \left(3H_4(y_e)H_3(y_\nu)+H_4(y_\nu)H_3(y_e)\right) \, , \nonumber\\
Q_{\scat}^{T}&=\frac{T^9 F_{e^-}}{18\pi^5\Lambda_{T}^4} \left(11H_4(y_e)H_3(y_\nu)+17H_4(y_\nu)H_3(y_e)\right) \, .
\label{Qfin}
\end{align}
Other channels like $e^-\bar\nu\to e^-\bar \chi$ or $e^+\nu\to e^+ \chi$ are obtained from Eq.~\eqref{eq:Q0_scatt} by replacing the arguments of the $H_n(y)$ functions accordingly. However, the scattering contributions in Eq.~\eqref{Qfin} give the dominant contributions to the total energy-loss rate, and result in the final lower bounds on the UV scales in the free-streaming regime, given by 
\begin{align}
\Lambda_X  \gtrsim \begin{cases}  \left(15  - 22\right)  \TeV  & X = V,A \\ \left(9.3 - 14 \right)\TeV \, & X = S,P \\ \left(25 - 37 \right)\,  \TeV & X = T \end{cases}  \, .
\end{align}
where the lower (upper) value corresponds to the SFHo18.80 (SFHo20.0) simulation.

\begin{align}
K^{X}_{s,{\scat}} &= \frac{2 s^2}{3\pi\Lambda_X^4}d_XF_eF_{\nu}  \, ,\\
 \Gamma^{X}_{s,{\scat}} & = \frac{4E_{\chi}T^4H_3(y_e)}{9\, \mathfrak{g}_{\chi}\pi^3\Lambda_X^4}d_XF_eF_{\nu}  \, ,\\
 \mathcal{ C}^{X}_{s,{\scat}} & = \frac{2T^8H_3(y_\nu)H_3(y_e)}{9\, \pi^5\Lambda_X^4}d_XF_eF_{\nu}  \, ,
 \label{Cscat}
\end{align}
with $d_{V,A} = 8\, d_{S,P} = d_T/7 = 1$. The trapping bounds are
\begin{align}
\Lambda_X  \gtrsim \begin{cases}  \left( 56 - 59  \right) \GeV & X = V,A \\ \left( 38 - 41 \right) \GeV  & X = S,P \\ \left( 88 - 92 \right) \GeV  & X = T \end{cases}  \, ,
\end{align}
covering the range obtained using the simulations SFHo18.80 or SFHo20.0.

\subsection{Effective operators with sterile neutrinos}
\label{sec:EFT_nuR}

\begin{figure*}[t!]
\begin{tabular}{cc}
    \centering    \includegraphics[width=0.5\columnwidth]{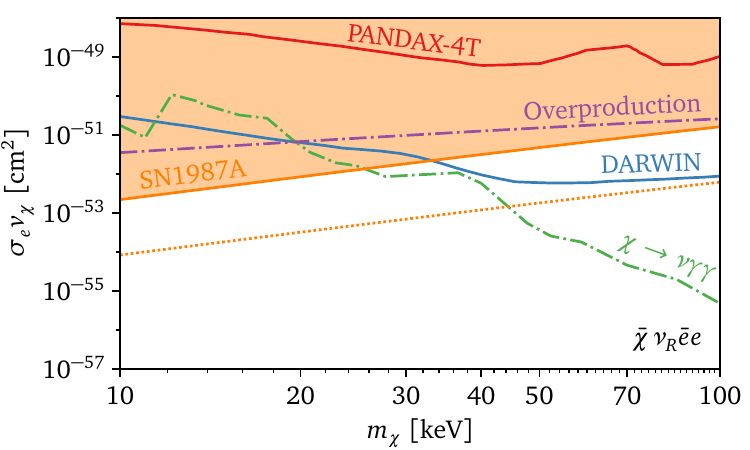} &
\includegraphics[width=0.5\columnwidth]{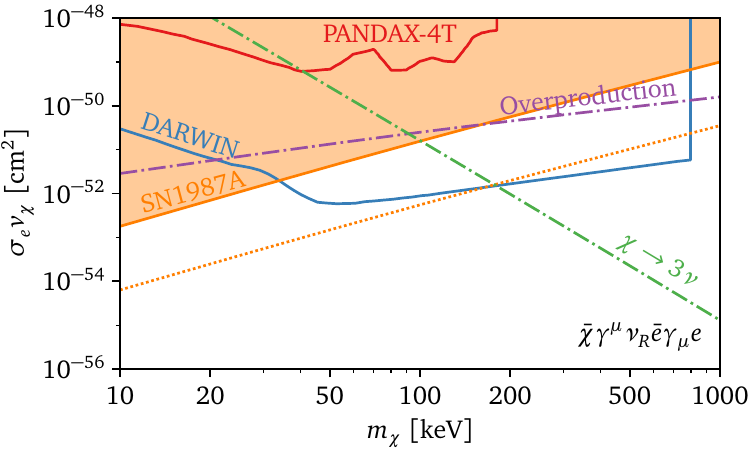}
\end{tabular}    
    \caption{SN~1987A cooling bound for scalar (left) and vector (right) interactions and sterile neutrinos. See caption of  Fig.~\ref{fig:money}. 
    \label{fig:moneyR}}
\end{figure*}

Another possibility that has been considered for absorption signatures in DD is to replace the SM neutrino by a sterile one in the effective Lagrangian~\eqref{eq:EFT_Lag} with $m_\nu\ll m_\chi$~\cite{Dror:2020czw} . In the context of the DM production in SNe, these models are substantially different from the ones considered with SM neutrinos. In the former case, the most important production mechanism is scattering $\nu_L e^-\to\chi e^-$, where SM neutrinos are in thermal equilibrium with the plasma. However, sterile neutrinos are not in equilibrium  and the only possible production mechanism is annihilation, $e^- e^+\to \chi\bar \nu_R$ and $e^- e^+\to \bar\chi\nu_R$. In fact, this is similar to the production of DM in SN with $\left(\bar ee\right)\left(\bar \chi\chi\right)$ interactions studied in Ref.~\cite{Manzari:2023gkt}, although in this case the emission of both $\chi$ and $\nu_R$ contributes to cooling.  

In Fig.~\ref{fig:moneyR}, we present the SN upper limits on DD cross sections, using the same style as in Fig.~\ref{fig:money}. These bounds are significantly weaker than those obtained for SM neutrinos, but are still complementary to both ID and DD. However, some parameter space remains accessible to DD searches for vector interactions, particularly within the region where the SN constraints are subject to uncertainties.

\section{Light Mediators}

We will investigate here the extension of the SN~1987A limits that we have derived in the EFT limit to the case where $(\overline \nu \chi )(\overline ee)$ interactions are produced by \textit{light} mediators, with mass $M\lesssim100$ MeV. In this case, mediators can be produced on-shell (resonantly) enhancing the efficiency of the dark production and absorption processes. For simplicity, in the following we focus on the massless approximations for electrons and dark particles $\chi$. Let us discuss the main contributions to the production and absorption resonant processes.

\subsection{Annihilation}

Resonant production of $\chi$ through $e^+ e^-\to X^*\to\chi\bar\nu$ occurs as long as the collision energies of the positrons and electrons in the plasma are sufficiently large, i.e. $m_{X}\lesssim 3T$. In this regime and taking the limit $m_{\chi} = m_e \to 0$, we obtain
\begin{align}
J_s=2\, C_{X} s^2 \frac{(E_1+E_2) g_e^2 \text{BR}_\chi}{m_X}F_{\bar\nu}\frac{\Gamma_X}{(s-m_X^2)^2+m_X^2\Gamma_X^2},
\label{defAC}
\end{align}
where $C_V = C_A = 2\, C_S = 2\, C_P = 1$, $\text{BR}_\chi=\Gamma(X\to\chi\bar\nu)/\Gamma_X$ is the branching fraction of $X$ into the dark channel $\chi\bar\nu$, and $\Gamma(X\to\chi\bar\nu)=A_X g_\chi^2m_X/12\pi $, with $A_V = 2\,A_S/3 = 1$, and $g_i$ the couplings of the mediator to fermion $i$. Assuming $\Gamma_X/m_X\ll1$ one can apply the narrow-width approximation (NWA), and obtain
\begin{align}
J_s=2\, C_X \pi m_X^2 (E_1+E_2) g_e^2 \text{BR}_\chi F_{\bar\nu}\delta(s-m_X^2),
\end{align}
and
\begin{align}
Q=C_X\frac{g_e^2 m_X^2\text{BR}_\chi T^3}{32\pi^3}F_{\bar\nu}\left(H_1(y_e)H_0(-y_e)+H_1(-y_e)H_0(y_e)\right).
\end{align}
One can also explicitly solve the full phase-space integrals with the resonant structure (beyond NWA) finding a good description of the production process below $m_X\simeq 100$ MeV for the benchmark SN conditions.

Likewise, one can adapt the calculation of the absorption width and rates of the process $\chi\bar\nu\to X^*\to e^+ e^-$ to the light-mediator regime,
\begin{align}
K_s=4\, C_X\, s^2 \frac{ g_e^2 \text{BR}_\chi}{m_X}F_eF_{\bar e}\frac{\Gamma_V}{(s-m_X^2)^2+m_X^2\Gamma_X^2}\,,
\end{align}
and in the NWA,
\begin{align}
K_s=4\, C_X\,\pi m_X^2 g_e^2 \text{BR}_\chi F_e F_{\bar e}\delta(s-m_X^2)\,.
\end{align}
In this limit we obtain,
\begin{align}
\Gamma_{\ann}(E_\chi)= C_X\, \frac{g_e^2m_X^2 T}{8\pi E_\chi^2\mathfrak g_\chi}\text{BR}_\chi F_e F_{\bar e} \log\left(1+e^{-y_{\nu}}\right),
\end{align}
and
\begin{align}
\mathcal C_{\rm abs}= C_X\, \frac{g_e^2m_X^2 T^2}{16 \pi^3}\text{BR}_\chi F_e F_{\bar e} \log\left(1+e^{-y_{ \nu}}\right) \log\left(1+e^{y_{\nu}}\right).
\end{align}

\subsection{Scattering}

Scattering processes like $e^-\nu\to e^-\chi$ arise by  $t$-channel exchange of the $X$ mediator. Although this process does not suffer from the suppression of the positron abundance in the plasma (see discussion in the EFT), it scales with the coupling constant as $\propto g^4$, compared to the $\propto g^2$ scaling of resonant annihilation in the light mediator regime. Thus, it is expected that for the small couplings required by the SN~1987A bound, the latter will dominate the emission and absorption rates. In this regime, for a vector mediator, we obtain
\begin{align}
J_s^{A,V}&=\frac{g_e^2g_\chi^2}{4\pi s^2}F_{e^-}\bigg(2 \log \left(\frac{s}{m_X^2}+1\right) \left(3 m_X^4 (E_1-E_2)+2 m_X^2 s (2 E_1-3 E_2)+2 s^2 (E_1-2
   E_2)\right)\nonumber\\
   &-\frac{E_1 s \left(6 m_X^4+11 m_X^2 s+7 s^2\right)}{m_X^2+s}+E_2 s \left(\frac{4 s^2}{m_X^2}+6 m_X^2+9
   s\right)\bigg).
\end{align}
while for a scalar mediator,
\begin{align}
J_s^{S,P} =\frac{g_e^2g_\chi^2}{16\pi s^2}F_{e^-}\bigg(2 \log \left(\frac{s}{m_S^2}+1\right) \left(3 m_S^4 (E_1-E_2) - 2 E_2 s\, m_S^2\right) + E_1 s \left(-4 m_S^2 + s - \frac{2 m_S^4}{m_S^2+s}\right)+E_2 s(s+6 m_S^2)\bigg).
\end{align}
A significant simplification can be achieved for light mediators expanding the previous formulae in $m_V$ and $m_S$ around zero, 
\begin{align}
J_s^{V,A}=\frac{g_\chi^2 g_e^2E_2 s}{\pi m_V^2}F_{e^-}\,, \qquad J_s^{S,P}=\frac{g_\chi^2 g_e^2(E_1+E_2)}{16\pi}F_{e^-}\,.
\end{align}
In this limit, one finally gets
\begin{align}
Q^{V,A}=\frac{g_\chi^2g_e^2 T^7}{8\pi^5m_V^2}F_{e^-}H_2(y_e)H_3(y_\nu)\,, \qquad Q^{S,P}=\frac{g_\chi^2g_e^2 T^5}{256\pi^5}F_{e^-}(H_2(y_e)H_1(y_\nu)+ H_1(y_e)H_2(y_\nu)).
\end{align}

The calculation of the absorption rate for the inverse scattering process yields
\begin{align}
\begin{split}
K_s^{V,A}&=\frac{g_e^2g_\chi^2}{2\pi}F_e F_\nu\left(\frac{m_V^2}{m_V^2+s}+\frac{2 s}{m_V^2}+\frac{2 \left(m_V^2+s\right) \log
   \left(\frac{m_V^2}{m_V^2+s}\right)}{s}+1\right)\,,\\
K_s^{S,P}&=\frac{g_e^2g_\chi^2}{4\pi\, s}F_e F_\nu\left(\frac{2\, s\, m_S^2 + s^2}{m_S^2+s}+2 m_S^2\log\left(\frac{m_S^2}{m_S^2+s}\right)\right)\,.
\end{split}
\end{align}
In the limit $m_V, m_S\to 0$ one obtains $K_s^{V,A}=g_e^2 g_\chi^2sF_e F_\nu/\pi m_V^2$, $K_s^{S,P}=g_e^2 g_\chi^2 F_e F_\nu/(4\pi)$ and
\begin{align}
\begin{split}
\Gamma_{\scat}^{V,A}(E_\chi) &=\frac{g_e^2 g_\chi^2 T^3}{4\pi^3 m_V^2 \mathfrak g_\chi}F_e F_\nu H_2(y_e),\hspace{0.6cm} \mathcal C_{\scat}^{V,A}=\frac{g_e^2 g_\chi^2 T^6}{8\pi^5 m_V^2}F_e F_\nu H_2(y_e)H_2(y_\nu)\,,\\
\Gamma_{\scat}^{S,P}(E_\chi) &=\frac{g_e^2 g_\chi^2 T^2}{32\pi^3E_{\chi} \mathfrak g_\chi}F_e F_\nu H_1(y_e),\hspace{0.6cm} \mathcal C_{\scat}^{S,P}=\frac{g_e^2 g_\chi^2 T^4}{64\pi^5}F_e F_\nu H_1(y_e)H_1(y_\nu).
\end{split}
\end{align}

\subsection{Photoproduction}
Photoproduction off electrons, $\gamma e^-\to V^*(\to\chi\bar \nu)e^-$ is negligible compared to annihilation and scattering for heavy mediators but needs to be considered for light mediators~\cite{Manzari:2023gkt}. We start with the volume emission rate, that will be approximated in the relativistic limit of electrons as~\cite{Manzari:2023gkt},
\begin{align}
Q_{\gamma}&=\frac{F_{e^-}}{12\pi^4}\int^\infty_0 d\omega \omega f_\gamma \int^\infty_0d\omega^\prime\omega^\prime (\omega+\omega^\prime)f_e\int^{+1}_{-1}d(\cos\theta)s\sigma(s)\, ,
\end{align}
where $\omega$ ($\omega^{\prime}$) is the photon (electron) energy, $\sigma(s)$ is the photoproduction cross section, $s=2\omega\omega^\prime(1-\cos\theta)$ and we have assumed that the $\chi$ in the process carries away $\approx1/3$ of the total energy in the collision. Given that this process is relevant only for light mediator masses, where the mediator $X$ is produced resonantly, we use the NWA.

For a vector mediator the proper cross section for $e^-\gamma\to V e^-$ was obtained in Eq.~(39) of Ref.~\cite{Manzari:2023gkt}. For the scalar mediator in Eq.~\eqref{eq:Lag_scalar_med} one obtains
\begin{align}
\sigma (\gamma e \to \phi e) & = \frac{s^2 e^2 y_{e}^2}{16 \pi  \left(s - m_e^2\right)^3} \left[ \left(1 - \frac{2 m_\phi^2}{s}+ \frac{6 m_e^2}{s}
    +  \frac{9 m_e^4}{s^2} -\frac{10 m_\phi^2 m_e^2}{s^2}+ \frac{2 m_\phi^4}{s^2} \right) \log
   \left(\frac{1 + (m_e^2-m_\phi^2)/s +\beta_f}{1+ (m_e^2-m_\phi^2)/s-\beta_f}\right) \right. \nonumber \\
  & \left.   - \frac{\beta_f}{2}
   \left(3 - \frac{7
   m_\phi^2} {s} + \frac{25 m_e^2}{s} + \frac{5 m_e^4}{s^2} - \frac{2 m_e^2 m_\phi^2}{s^2} + \frac{m_e^4 m_\phi^2}{s^3} -\frac{m_e^6}{s^3} \right) \right]
\end{align}
with $\beta_f = \sqrt{(1 - m_\phi^2/s)^2 -2 m_e^2/s (1 + m_\phi^2/s) + m_e^4/s^2}$.

\subsection{Application to a simplified vector model coupled to sterile neutrinos}

\begin{figure}[t!]
    \centering    \includegraphics[width=0.6\linewidth]{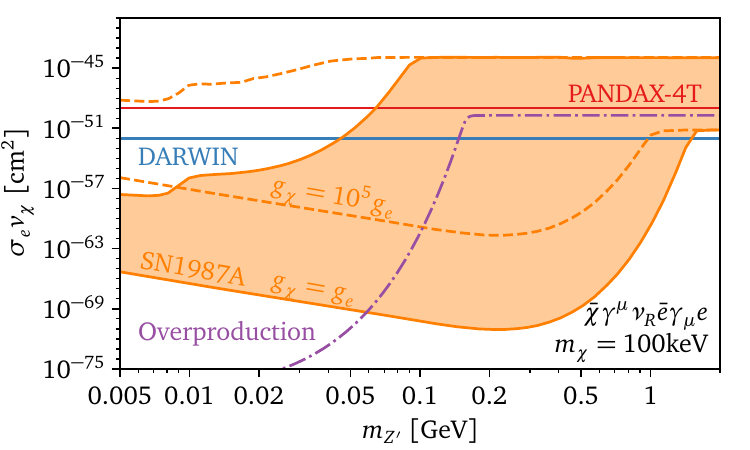}
    \caption{Bounds on the DD cross-section for the absorption of DM with $m_\chi=100$ keV, a sub-GeV vector mediator and a sterile neutrino. See caption of Fig.~\ref{fig:light_scalar}.
}
    \label{fig:light_vector}
\end{figure}

In the main article we discussed the importance of considering carefully the light-mediator regime when studying the SN bounds on sub-MeV DM models. The reason is the onset of resonant processes that change the dependence of the bounds in the couplings. 
We studied this for the scalar $(\bar e e)(\bar\nu\chi)$ interactions in Fig.~\ref{fig:light_scalar}. Here, we repeat the analysis for a simplified vectorial model featuring interactions with a sterile neutrino (instead of a SM one)~\cite{Dror:2020czw},
\begin{equation}
\mathcal L_V\supset \left(g_\chi\bar\chi \gamma^\mu \nu_R +g_e\bar e \gamma^\mu e\right)Z^\prime_\mu.
\label{eq:Lag_vector_med}
\end{equation}
As discussed above, in case of sterile neutrinos the calculations are analogous to those for the $\left(\bar ee\right)\left(\bar \chi\chi\right)$ interactions studied in Ref.~\cite{Manzari:2023gkt}. Following this reference, in Fig.~\ref{fig:light_vector} we show the dependence of the bounds as a function of the mediator mass $m_{Z^\prime}$ for a DM mass $m_\chi=100$ keV. 
As in the scalar case represented in Fig.~\ref{fig:light_scalar}, the SN bounds become much stronger for light $Z^\prime$ opening up a region of cross sections that could be probed by DD. However, as in the case of the scalar mediator, cosmological production also becomes resonant, further constraining the allowed parameter space.

\section{Overproduction Limit}
 In this section we discuss  the limits from DM overproduction in both the EFT regime (the operators in Eq.~\eqref{eq:EFTLag}) and the scalar and vector UV completions in Eq.~\eqref{eq:Lag_scalar_med} and Eq.~\eqref{eq:Lag_vector_med}, respectively.  We follow Ref.~\cite{Lehmann:2020lcv} and integrate the Boltzmann equation in the freeze-in regime~\cite{Hall:2009bx}, see the appendices in Refs.~\cite{DEramo:2017ecx, Badziak:2024szg} for details. For the effective operators, the freeze-in contribution to the DM abundance is UV dominated, so that the abundance will grow with the reheating temperature $T_R$ and decrease with the EFT scale $\Lambda$. In order to derive conservative  lower bounds on $\Lambda$ from DM overproduction, we thus assume the lowest possible reheating temperature $T_R = 4 \MeV$~\cite{Kawasaki:2000en, Hannestad:2004px} and take a  vanishing initial DM abundance (any non-zero abundance would make the limit stronger).

\subsection{General Case}

For a general $2 \to 2$ process $\chi + 1 \to 2 + 3$ involving a single DM particle $\chi$ and particles $2$ and $3$ in the thermal bath, the  Boltzmann Equation (BE) for the DM  yield $Y_\chi \equiv n_\chi/s$ is given by
\begin{align}
\frac{d Y_{\chi}}{dT}  & = -  \frac{1} {s H T} \left( 1+ \frac{dg_{*s}}{dT}  \frac{T}{3g_{*s}} \right) \left[-  \langle \sigma_{\chi 1 \to 2 3 } v \rangle n_\chi n_1  + \langle \sigma_{23 \to \chi 1 } v \rangle n_2^{\rm eq} n_3^{\rm eq} \right] \nonumber \\
& = -  \frac{1} {s H T} \left( 1+ \frac{dg_{*s}}{dT}  \frac{T}{3g_{*s}} \right) \langle  \sigma_{\chi 1 \to 23 } v \rangle  \left[ n_\chi^{\rm eq} n_1^{\rm eq}  - n_\chi n_1 \right] \nonumber \\
& = -  \frac{1} {s H T} \left( 1+ \frac{dg_{*s}}{dT}  \frac{T}{3g_{*s}} \right)  {\cal C}_{\chi 123 }   \left[1 - Y_\chi Y_1/ (Y_\chi^{\rm eq} Y_1^{\rm eq})  \right]  \, , 
\end{align}
with the collision operator $  {\cal C}_{\chi 1  23 }   =  \langle  \sigma_{\chi 1 \to 23 } v \rangle   n_\chi^{\rm eq} n_1^{\rm eq} =    \langle  \sigma_{23 \to \chi 1} v \rangle   n_2^{\rm eq} n_3^{\rm eq}$. It is convenient to introduce a reference scale $m$ by defining $T = m/x$, so that the BE becomes
\begin{align}
\frac{d Y_{\chi}}{dx}  & =   \frac{1} {s H x} \left( 1- \frac{dg_{*s}}{dx}  \frac{x}{3g_{*s}} \right)  {\cal C}_{\chi 123}   \left[1 - Y_\chi Y_1/ (Y_\chi^{\rm eq} Y_1^{\rm eq}) \right]  \, .
\end{align}
Taking the effective degrees of freedom $g_*$ to be approximately constant, one has $s \propto T^3, H \propto T^2$, and the BE simplifies to 
\begin{align}
\frac{d Y_{\chi} (x)}{dx}  & =   \frac{ x^4 {\cal C}_{\chi 1  23 } (x)} {s (m) H (m) }  \left[1 - Y_\chi (x) Y_1 (x)  / (Y_\chi^{\rm eq} (x) Y_1^{\rm eq} (x)) \right] \, .
\label{BE}
\end{align}
In the present scenario  it is clear that DM production continues at most until $T \approx m_e$, when the electrons become non-relativistic, so that $Y_1 (x) \lesssim Y_1^{\rm eq (x)}$ for $1 = \nu, e$. Moreover in the freeze-in limit the DM abundance never reaches equilibrium, $Y_\chi \ll Y_\chi^{\rm eq}$, so that one can neglect the whole bracket on the RHS in Eq.~\eqref{BE}  to good approximation.  We can then integrate both sides easily from the maximal temperature $T_R$ with vanishing yield down to  $T = 0$ for the asymptotic yield.  More appropriate would be to integrate until the temperature where both electrons and neutrinos $T \approx 1 \MeV$, where both electrons and  neutrinos have decoupled from the thermal path and electrons  start to become non-relativistic, but in practice this difference does not matter much, as the integral is UV dominated and the IR boundary gives only a small correction. We thus obtain for the asymptotic yield $Y_{\infty}$
\begin{align}
{Y_{\infty}} & \approx    \int_{m/T_R}^\infty   \frac{x^4 {\cal C}_{\chi 1  23} (x) } {s (m) H (m) } \, dx=\frac{135 \sqrt{5} M_{\rm Pl}}{4 m^5 \pi^{7/2}  g_*^{3/2} (m) }   \int_{m/T_R}^\infty x^4 {\cal C}_{\chi 1  2 3 } (x) \, dx \equiv  R(m) \,  .
\label{Rgen}
\end{align}
 Including a factor 2 for the abundance for $Y_{\overline{\chi}}$  we finally get for the total relic abundance
\begin{align}
\Omega_\chi  h^2 & \approx 2 R(m) m_\chi  s_0/\rho_{\rm crit} = 0.12 \left( \frac{m_\chi}{10 \keV} \right) \left( \frac{R(m)}{2.2 \times 10^{-5}} \right)  \, .
\label{Omh2gen}
 \end{align} 

\subsection{Effective Operators}
To calculate the relic abundance for the EFT operators in Eq.~\eqref{eq:EFTLag}, we merely need to integrate the collision term. For simplicity we will work in the high-energy limit and neglect $m_e$ and $m_\chi$, in which case the collision terms for annihilation $e^+ e^- \to \chi \overline \nu$ and scattering  $e^\pm \nu \to e^\pm \chi $ are given in Eq.~\eqref{Cann} and Eq.~\eqref{Cscat}, respectively, in the limit of vanishing chemical potentials. Also taking the Boltzmannian limit for simplicity, we obtain
\begin{align}
{\hat  {\mathcal C}}_{e^+ e^- \to \chi \overline \nu} & = \frac{4 a_X T^8}{\pi^5 \Lambda_X^4} \, , & 
{\hat  {\mathcal C}}_{e^\pm \nu \to e^\pm \chi} & = \frac{8 d_X  T^8}{\pi^5 \Lambda_X^4} \, ,
\end{align}
where $a_X$ and $d_X$ are given below Eq.~\eqref{Jhatsann} and  Eq.~\eqref{Cscat}, respectively. The collision operator can be easily integrated, giving for annihilation 
\begin{align}
  \int_{m/T_R}^\infty x^4 {\cal C}_{e^+ e^- \to \chi \overline \nu } (x)  \, dx=  \frac{4 a_X  m^8 }{\pi^5 \Lambda_X^4}  \int_{m/T_R}^\infty \frac{dx}{x^4} = \frac{4 a_X m^5 T_R^3 }{3 \pi^5 \Lambda_X^4}  \, , 
  \end{align}
  and similar for scattering with $a_X \to 2 d_X$ (note that for RH neutrinos there is only annihilation, provided that there are not in thermal equilibrium in the early universe).  Summing the contributions   from annihilation and scattering (including a factor of 2 for charge multiplicity), we obtain 
 
  \begin{align}
 R(m) &  =    \frac{45 \sqrt{5} M_{\rm Pl}}{  \pi^{17/2}  g_*^{3/2} (m) }   \left( a_X  + 4 d_X \right) \frac{  T_R^3 }{  \Lambda_X^4}   \, , 
  \end{align}
and finally
\begin{align}
\Omega_\chi  h^2 & \approx 0.12 \, \left( a_X  + 4 d_X \right)  \left( \frac{m_\chi}{10 \keV} \right) \left( \frac{T_R}{4 \MeV} \right)^3  \left( \frac{1.6 \TeV}{\Lambda_X} \right)^4 \left( \frac{10.7}{g_* (m)} \right)^{3/2} \, , 
 \end{align} 
where we normalized to $g_*(T_R =4 \MeV) = 10.7$, since this regime gives the dominant contribution to the abundance.  Doing the integral of the collision term with the full cross-section, i.e. keeping finite $m_\chi = m_e \ne 0$, results in a scale that differs from the approximation above only by a few GeV. Directly solving the BE numerically with the full cross-section gives a difference of the same size, which demonstrates that the choice of $m= T_R$ is indeed  appropriate. As discussed above we have neglected here the IR boundary term, which would change the result to $T_R^3 \to T^3 - T_{\rm IR}^3$. For $T_{\rm IR} \approx 1 \MeV$ and our default value $T_R = 4 \MeV$, this would merely amount to a 2\% correction. Finally fixing $T_R = 4 \MeV$, the limit from DM overproduction  reads for scalar and vector operators (valid for $m_\chi \ll T_R$)
\begin{align}
 \Lambda_{S} & \gtrsim 1.7 \TeV \left( \frac{m_\chi}{10 \keV} \right)^{1/4} \, , & 
  \Lambda_{V} & \gtrsim 2.3 \TeV \left( \frac{m_\chi}{10 \keV} \right)^{1/4}
 \end{align} 
or 
\begin{align}
\frac{m_\chi^2}{4 \pi \Lambda_{S}^4} & \lesssim 4.1 \times 10^{-52} {\rm cm}^2 \left( \frac{m_\chi}{10 \keV} \right) \, , & 
\frac{m_\chi^2}{4 \pi \Lambda_{V}^4} & \lesssim 1.0 \times 10^{-52} {\rm cm}^2 \left( \frac{m_\chi}{10 \keV} \right) \, ,
\end{align} 
in good agreement with Fig.~\ref{fig:money}.

\subsection{UV Completions}

For a light vector or scalar mediator $X$, in principle we would need to solve two coupled Boltzmann Equations for $\chi$ and $X$. However,  since we are interested in the limit $T_R \ll m_V $ and unstable mediators, the mediator abundance is always negligible and we can treat $X$ as a short-lived resonance and simply work with a single Boltzmann equation for $\chi$ (c.f. also Ref.~\cite{Frangipane:2021rtf}). Thus we can employ again the general result from Eq.~\eqref{Rgen}, using the  collision term for the relevant process in the UV complete theory.

Since we  expect significant deviations from the EFT calculation only in the resonant regime (i.e. when the mediator is produced on-shell) we restrict to annihilation.  The corresponding s-channel cross-section for a vector mediator reads
\begin{align}
\sigma_{e^+ e^- \to \chi \overline \nu} & =  \frac{g_e^2 g_\chi^2 \sqrt{1- 4 m_e^2/s} }{48\pi \left( (s - m_V^2)^2 + m_V^2 \Gamma_V^2  \right)s^2(s-4m_e^2)} (s+2m_e^2)(s-m_\chi^2)^2(2s+m_\chi^2)\,,
\end{align}
which results in the collision term (taking $m_\chi \ll m_e$)
\begin{align}
{\cal C}_{e^+ e^- \to \chi \overline \nu} & = \frac{T}{8 \pi^4} \int_{4 m_e^2}^\infty \left( 1- 4 m_e^2/s \right) s^{3/2} \sigma_{e^+ e^- \to \chi \overline \nu} K_1 \left( \sqrt{s} /T \right) \, ds \, ,
\end{align}
and the partial widths are given by
\begin{align}
\Gamma (V \to e \overline e) & = \frac{ g_e^2 m_V}{12 \pi } \left( 1 + 2 \frac{m_e^2}{m_V^2} \right) \sqrt{1 - 4 \frac{m_e^2}{m_V^2}} \, , \nonumber \\
\Gamma (V \to \chi \overline \nu) & = \frac{ g_\chi^2 m_V}{12 \pi } \left( 1 +  \frac{m_\chi^2}{2m_V^2} \right) \left( 1 -  \frac{m_\chi^2}{m_V^2} \right)^2 \, .
\end{align}
In the resonant limit that we are interested in, the dominant contributions come from $\sqrt s \approx m_V  \gg m_e \gg m_\chi$. Therefore, we  now set $m_\chi = m_e = 0$ and   use the NWA for the cross-section, giving 
\begin{align}
\sigma_{e^+ e^- \to \chi \overline \nu}  &   \approx g_e^2 \pi   \delta (s - m_V^2)  {\rm BR}  (V \to \chi \overline \nu)  \, , 
\end{align}
and thus
\begin{align}
{\cal C}_{e^+ e^- \to \chi \overline \nu } & = \frac{g_e^2 }{8 \pi^3} m_V^3 T  K_1 \left( \frac{m_V}{T} \right)   {\rm BR}  (V \to \chi \overline \nu) \, .
\end{align}
Therefore the final integral becomes
\begin{align}
\int_{m/T_R}^{\infty} x^4 {\cal C}_{e^+ e^- \to \chi \overline \nu } (x) = \frac{g_e^2 m_V^3 m}{8 \pi^3}  {\rm BR}  (V \to \chi \overline \nu) \int_{m/T_R}^{\infty} x^3   K_1 \left( \frac{m_V }{m} x \right)  \, .
\end{align}
Now choosing $m = m_V$, and taking $m_V \gg T_R$, it is clear that the integral is again UV dominated, which justifies to integrate to $T=0$ instead to some finite value. This gives for the asymptotic yield

 \begin{align}
R &   = \frac{135 \sqrt{5} M_{\rm Pl} }{32 m_V \pi^{13/2}  g_*^{3/2} (m_V) }   \frac{g_e^2 g_\chi^2}{g_e^2 + g_\chi^2} \int_{m_V/T_R}^{\infty} x^3 K_1 (x) \, dx \, , 
\end{align}
and finally for the relic abundance
\begin{align}
\Omega_\chi  h^2 & = 0.12 \left( \frac{g_e g_\chi/\sqrt{g_e^2 + g_\chi^2}}{4.9 \times 10^{-12}} \right)^2  \left( \frac{m_\chi}{10 \keV} \right)  \left( \frac{10 \MeV}{m_V} \right)  \left( \frac{10.7}{g_*(m_V)} \right)^{3/2}   K \, ,  
 \end{align} 
 with 
 \begin{align}
 K & =  \frac{2}{3 \pi} \int_{m_V/T_R}^{\infty} x^3 K_1 (x) \, dx  \, , 
 \end{align}
 so that for $T_R \to \infty$, $K\to 1$, so that one recovers the usual IR freeze-in result from the decay of a  particle $V$ in the thermal bath~\cite{Hall:2009bx}. Instead  the finite reheating temperature result in additional suppression for increasing mediator masses, for example
\begin{align}
\left( \frac{10 \MeV}{m_V} \right)  \left( \frac{10.7}{g_*(m_V)} \right)^{3/2}   K & = \begin{cases}
6.3 \times 10^{-1} & m_V = 10 \MeV \\
8.7 \times 10^{-5} & m_V = 50 \MeV \\
6.1 \times 10^{-10} & m_V = 100 \MeV 
\end{cases} \, .
\end{align}
Finally it is convenient to introduce
$g_\chi  \equiv \kappa g_e$, $\Lambda_V  = m_V/(g_e \sqrt \kappa)$
so that
\begin{align}
\Omega_\chi  h^2 & = 0.12  \, C_X A_X  \left( \frac{2 \kappa}{1+\kappa^2} \right) \left( \frac{1.4 \times 10^9  \GeV}{ \Lambda_X} \right)^2  \left( \frac{m_\chi}{10 \keV} \right)  \left( \frac{m_X}{10 \MeV} \right)  \left( \frac{10.7}{g_*(m_X)} \right)^{3/2}   K \, ,  
 \end{align} 
 where we have generalized our result to include scalar mediators with $C_X$ and $A_X$ defined below Eq.~\eqref{defAC}.
Comparing to the limits in the EFT regime from the previous section, one can estimate the transition region by equating the abundances and solving for $m_V$, which gives for $m_\chi = 10 \keV, T_R = 4 \MeV$ the value $m_X \approx 160 \MeV$, in good agreement with Fig.~\ref{fig:light_scalar} and \ref{fig:light_vector}.

\end{document}